\documentclass[twocolumn,aps,prb]{revtex4-2}

\usepackage{graphicx}
\usepackage{bm}
\usepackage{float}
\usepackage{amsthm,amssymb,amsmath,url,braket,amsbsy} %%%MATH
\usepackage{threeparttable}
\usepackage[utf8]{inputenc}
\usepackage{dirtytalk}
\graphicspath {{./Figure/}}
\usepackage{color}
\usepackage{cancel}

\usepackage{tikz} 
\usetikzlibrary{shapes,arrows,positioning,automata,backgrounds,calc,er,patterns}
\usepackage[compat=1.1.0]{tikz-feynman}

\begin{document}

\title{ Intrinsic Orbital and Spin Hall Effect in Bismuth Semimetal }

\author{Guanxiong Qu}
\affiliation{RIKEN Center for Emergent Matter Science (CEMS), Wako 351-0198, Japan.}
\author{Gen Tatara}
\affiliation{RIKEN Center for Emergent Matter Science (CEMS), Wako 351-0198, Japan.}
%\affiliation{National Institute for Materials Science, Tsukuba 305-0047, Japan}

\date{\today}

\begin{abstract}
We investigate the intrinsic orbital Hall conductivity (OHC) and spin Hall conductivity (SHC)  in Bismuth semimetal, by employing $sp$-orbital tight-binding model. We report a notable difference between anisotropy of OHC and SHC whose orbital and spin polarization lie within the basal plane. %We report a disparate anisotropy of OHC and SHC whose orbital (spin) polarization within the basal. 
We do not observe substantial correlation between the orbital and spin Berry curvature with spin orbit coupling (SOC) at Fermi energy, disproving the correlation between OHC and SHC in strong SOC regime. We argue the huge SHC in Bi semimetal attributes to its gigantic SOC which strongly affects the hybridization of the $p$-orbitals, despite its relatively small magnitude of OHC. We hope the distinct anisotropy of OHC and SHC provides a feasible mean to differentiate the two effects in experiments.
\end{abstract}

%\keywords{}

\maketitle

\date{\today}
\section{Introduction} \label{Sec1: Intro}
The spin Hall effect (SHE) \cite{hirsch1999spin,murakami2003dissipationless,sinova2015spin} is a phenomenon in which applying an external electric field generates a transverse spin current and finally leads spin magnetic moment accumulation at edges \cite{kato2004observation}. The SHE is considered as an essential approach for charge to spin conversion with which the field of spintronics flourishes. The orbital Hall effect (OHE) \cite{bernevig2005orbitronics} is a natural extension of SHE, claiming that a transverse orbital current can also be induced by external electric field and brings about the field of orbitronics \cite{go2021orbitronics}. The similarity of OHE and SHE stimulates people to investigate their correlation where people has claimed that SHE is converted from OHE via SOC \cite{go2018intrinsic}. Still, a clear distinction between the two sources of magnetic moment flow is lack of experimental proof \cite{go2021orbitronics}, leaving the relationship between OHE and SHE an open debate.

Historically, p-orbital semiconductors such as Si and Ge have been early candidates for investigations into spin Hall effect (SHE) and orbital Hall effect (OHE). However, both theoretical predictions \cite{bernevig2005orbitronics,guo2005ab,PhysRevLett.95.166605} and experimental results \cite{kato2004observation,ando2012observation} have shown relatively small magnitudes of SHE and OHE on these materials. Recent experiments have indicated that Bismuth semimetal \cite{hou2012interface,PhysRevB.93.174428} and its alloys with other p-orbital materials \cite{chi2020spin,PhysRevB.105.214419} exhibit relatively strong SHE ($\sigma_{SH} \sim 610$ $(\hbar/e) \Omega^{-1} cm^{-1}$) compared to other $d$-orbital heavy metals, suggesting Bismuth semimetal as a candidate for achieving large spin-charge conversion without the use of 5$d$ transition metals. Furthermore, Bi is the most commonly used element for forming topological insulators \cite{hasan2010colloquium}, whose unique surface state has attracted significant interest in solid-state physics, e.g., its potential for achieving high spin-conversion efficiency in spintronics \cite{mellnik2014spin,fan2014magnetization}. However, experiments indicate that spin-charge conversion dominantly arises from the intrinsic contribution of bulk states \cite{PhysRevB.105.214419,chi2020spin}, which is consistent with theoretical studies employing tight-binding calculations \cite{csahin2015tunable} and an effective model Dirac Hamiltonian \cite{fukazawa2017intrinsic, fuseya2012spin, PhysRevB.105.214419}. Additionally, the strong crystal field of the Bi rhombohedral structure induces a significant anisotropy in the spin Hall conductivity (SHC), as predicted by density functional theory (DFT) calculations \cite{guo2022anisotropic}, which has been supported by recent experiments using high-quality Bi films \cite{fukumoto2022observation,liang2022anisotropic}. However, theoretical investigations into OHE in Bismuth semimetal are currently lacking. Therefore, we believe that the combination of its enormous strength of spin-orbit coupling (SOC) and strong crystal field splitting make Bi semimetal an excellent candidate for investigating the relationship between OHE and SHE beyond the weak SOC regime \cite{go2018intrinsic}.

In this paper, we theoretically investigate the intrinsic OHE and SHE of Bismuth semimetal through tight-binding model and Berry phase formalism. We find a significant disparity between the anisotropy of OHC and SHC whose orbital and spin polarization lie within the basal plane. By comparing the orbital and spin Berry curvature, we argue the OHE and SHE do not exhibit substantial correlation in the strong SOC regime. We further show that the gigantic SOC in Bi yields a huge SHE via $p$-orbital hybridization but suppress its OHE.

\section{Model} \label{Sec2: Model}
Bismuth is a rhombohedral crystal of space group $R\bar{3}m$ and has two atoms per unit cell. It is a typical semimetal with hole pockets at T point and electron pockets at L points in the Brillouin zone. Notably, the direct gap at L point is extremely small around $13.6$ meV \cite{vecchi1974temperature} which is usually overestimated in calculations via DFT with generalized gradient approximation \cite{gonze1990first,guo2022anisotropic}. Nevertheless, Liu et al. purposed a third nearest-neighbor tight-binding model with SOC and successfully depicted the band structure of bismuth around the Fermi energy \cite{liu1995electronic}. Here, we employ such model constructed from $s$,$p$ orbitals of each atom with first, second, and third neighbor hopping parameters $V_{\alpha \beta},V'_{\alpha \beta},V''_{\alpha \beta}$ $(\alpha,\beta=s,p_x,p_y,p_z)$, shown in Fig.~\ref{fig:1} (b). Bi is featured by its gigantic SOC strength $\lambda=1.5$ $eV$ whose Hamiltonian in orbital basis reads \cite{PhysRevB.16.790}
\begin{align}
H_{so}=\frac{2\lambda}{3\hbar^2} \bm{L} \cdot \bm{S}
\label{eq:1}
\end{align}
where $\bm{L} =(L_x,L_y,L_z)$ are local orbital angular moment (OAM) operator of the $p$-orbitals and $\bm{S}=(S_x,S_y,S_z)$ are spin operator for each $s$,$p$ orbitals. 

The intrinsic OHC and SHC are calculated via Kubo formula in the clean limit \cite{guo2008intrinsic,tanaka2008intrinsic},
 \begin{align}
 \label{eq:2}
\sigma _{ji}^{X_k} &= \frac{ e}{ \hbar} \sum_{n, \bm{k}} f_{n,\bm{k}} \Omega^{X_k}_{ji,n} (\bm{k}) \\
\Omega^{X_k}_{ji,n}  &= \hbar \sum_{n \neq m} 2 \text{Im} \frac{ \bra{n} v^{X_k}_j \ket{m}  \bra{m} v_{i} \ket{n}  }{ ( \varepsilon_n - \varepsilon_m  )^2 }
\label{eq:3}
\end{align}
where $ f_{n,\bm{k}}$ is the Fermi distribution function, $\Omega^{X_k}_{ji,n}$ are spin $(X_k=S_k)$ and orbital $(X_k=L_k)$ Berry curvature of band $n$. Here, $ v^{X_k}_j = \{X_k, v_j\}/2$ are spin and orbital current with velocity operator defined as $v_j= \partial_{k_j} H/\hbar$. Note that OAM operator above is defined within each atomic site which is usually termed as atom-centered approximation (ACA) \cite{hanke2016role}. The modern theory of orbital magnetization \cite{shi2007quantum, PhysRevLett.95.137205} only offers the diagonal component of the OAM operator in Bloch basis, for which it can not be used to construct the orbital current.

\section{Results} \label{Sec3: results}
\begin{figure}[h]
 \includegraphics[width=8cm]{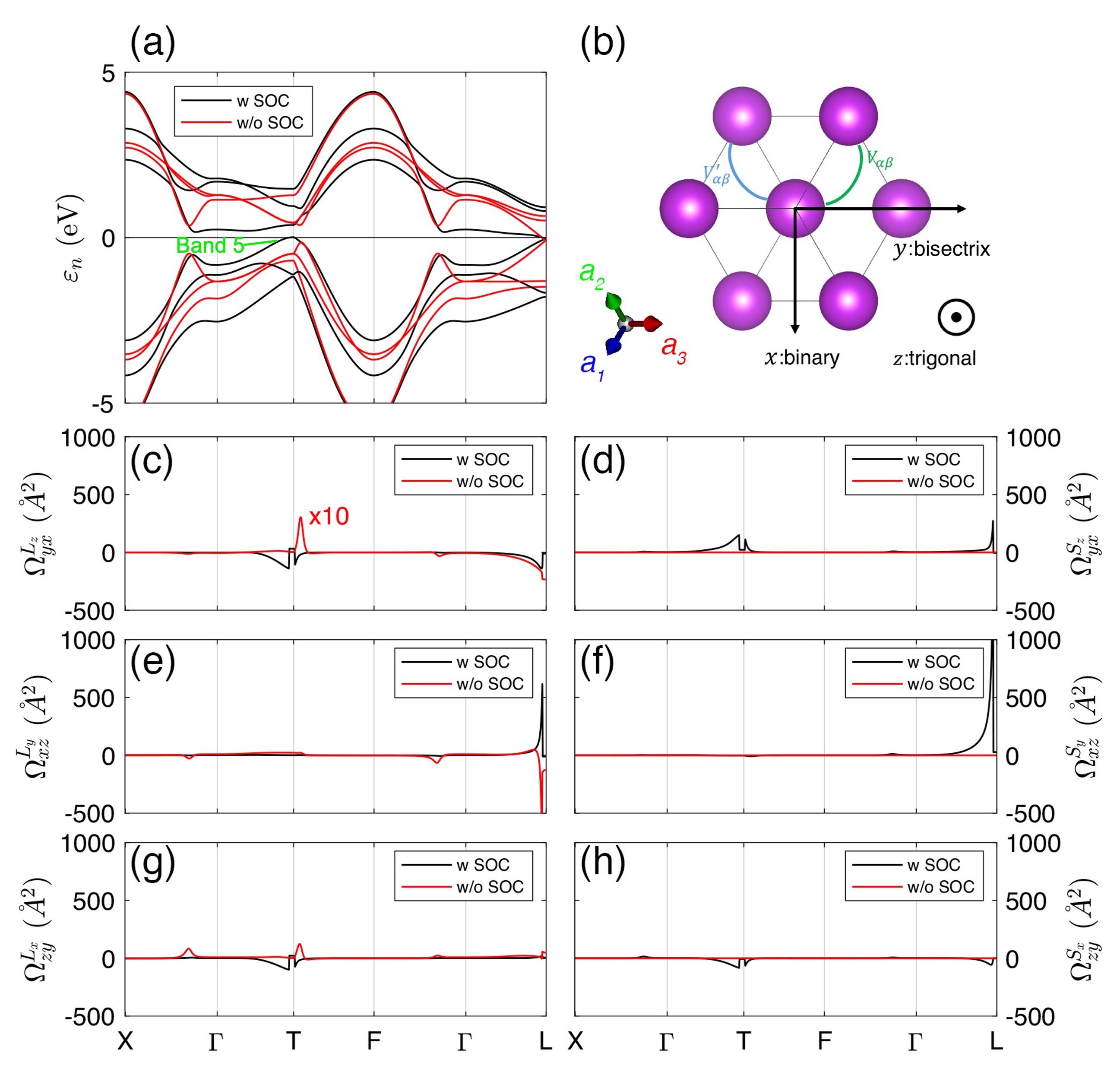}
 \caption{\label{fig:1} (a) Band structure, (c,e,g) total orbital Berry curvature, and (d,f,h) total spin Berry curvature of Bismuth with (black) and without (red) spin-orbit coupling. (b) the crystal structure of Bismuth show the $1st$ $(V_{\alpha \beta})$ and $2nd$ $(V'_{\alpha \beta})$ nearest neighbour hopping.}
 \end{figure}

The electronic band structure near the Fermi energy is shown in Fig.~\ref{fig:1} (a), which reproduces the previous tight-binding investigations \cite{liu1995electronic,csahin2015tunable} featured by the T-hole pockets and L-electron pockets. We emphasize the significant band modification induced by the gigantic spin-orbit coupling of Bi. For example, along the $\Gamma$-T line, without SOC the $p_z$-orbital dominated bonding state is below two degenerate $p_{x,y}$ bonding states. After inducing the SOC, the degeneracy of two $p_{x,y}$ states is lifted and one of the $p_{x,y}$ states hybridizes with $p_z$ orbital forming a gap at T point. A similar $p$-orbital hybridization and degeneracy lift induced by the SOC occur in the anti-bonding states of the conduction bands. The strong orbital hybridization induced by the gigantic SOC is essential to the SHE \cite{PhysRevB.99.064410}.

%%%%%%%%%% Fig:1 %%%%%%%%% 
 %%%%%%%%%%%  Fig:1 %%%%%%%%% 
 
The OHC and SHC tensors ($\sigma^{X_k}_{ji}$) generally have 27 components, most of which can be eliminated by the symmetry operations of the (magnetic) point group of the specific material investigated \cite{PhysRevB.92.155138}.  By such symmetry analysis, orbital and spin conductivity tensor of Bi has four non-vanishing components, $\sigma^{X_z}_{yx}, \sigma^{X_y}_{xz}, \sigma^{X_x}_{zy}$,and $\sigma^{X_x}_{yy}$ \cite{guo2022anisotropic}. We note that the tensor component  $\sigma^{X_x}_{yy}$ is not zero but component  $\sigma^{X_y}_{xx}$ vanishes as reported in Ref.~\cite{guo2022anisotropic}, indicating the disparity of binary and bisectrix axes [see Fig.~\ref{fig:1}(b)]. $\sigma^{L_x}_{yy}$ and  $\sigma^{S_x}_{yy}$ are $-102.9$ and $22.4$ $(\hbar/e) \Omega^{-1} cm^{-1}$, respectively, indicating the longitudinal spin current generation is relatively minor. In our following calculation, we only focus on the three conventional components, $\sigma^{X_z}_{yx}$, $\sigma^{X_y}_{xz}$, and $ \sigma^{X_x}_{zy}$.

We present the total orbital Berry curvature ($\Omega^{L_k}_{ji}$) and total spin Berry curvature ($\Omega^{S_k}_{ji}$) which are summations over all occupied bands ($\Omega^{X_k}_{ji} (\bm{k}) = \sum_n  f_{n,\bm{k}}\Omega^{X_k}_{ji,n} (\bm{k})$) along the high symmetry line in Fig.~\ref{fig:1}. For all the tensor components, the major contributions to OHE and SHC mainly concentrate on the T and L points where the hole and electron pockets locate, respectively. However, the three different geometries of total orbital (spin) Berry curvature exhibit distinct profiles, implying an anisotropic OHC (SHC) tensor. Without SOC, $\Omega^{S_k}_{ji}$ for all components vanish exactly, while $\Omega^{L_k}_{ji}$ are still finite and show dominant contributions around T and L points. Our investigation confirms that the OHE can exist in the absence of SOC, while SHE appears simultaneously with spin orbit interaction \cite{go2018intrinsic}.

\subsection{Anisotropy of OHC and SHC}
For a rhombohedral crystal, Bi semimetal is expected to exhibit anisotropic OHE and SHE. We use two parameter to characterize the anisotropy of OHC and SHC: $\Delta^{zy}_{X}= (\sigma^{X_z}_{yx}-\sigma^{X_y}_{xz} )/\sigma^{X_y}_{xz}$ and $\Delta^{yx}_{X}= (\sigma^{X_y}_{xz}-\sigma^{X_x}_{zy} )/\sigma^{X_x}_{zy} $ $(X=L,S)$, representing the anisotropy whose orbital and spin polarization are normal to the basal plane and within the basal plane [see Fig.~\ref{fig:1} (b)]. The anisotropy of SHC normal to the basal $\Delta^{zy}_{S}= 29.1\%$ is much larger than that within the basal $\Delta^{yx}_{S}=0.3\%$, semiquantitatively consistent with precious first principles calculations \cite{guo2022anisotropic}. We argue the quantitative discrepancy attributes to the detailed band structure between DFT and tight-binding calculations and the overestimation of the band gap of L point which is essential for evaluating the SHC and OHC \cite{guo2008intrinsic,tanaka2008intrinsic}. %However, the anisotropy of OHC is found to be comparatively larger than that of SHC (e.g., $\Delta^{zy}_{L}=107$ $(\hbar/e) \Omega^{-1} cm^{-1}$, dropped by about $38\%$), which even has a sign change within the basal plane, see Table~\ref{tab:1}. 
However, the anisotropy of OHC with in the basal plane is found to be comparatively larger than that of SHC $\Delta^{yx}_{L}=-245.7\%$, which even has a sign change within the basal plane, see Table~\ref{tab:1}. 
The discrepancy between anisotropy of OHC and SHC, particularly within the basal plane, is intriguing. Since the OHE is widely claimed as a more fundamental mechanism which induces SHE in the presence of SOC \cite{go2018intrinsic}, the strongly anisotropic OHE in the basal plane should not be expected to generate a nearly isotropic SHE, considering a isotropic SOC (Eq.~\ref{eq:1}). Therefore, in the following section we first investigate the cause of anisotropy of OHC and SHC in Bi semimetal.

%%%%%%%%%%%  Fig:2 %%%%%%%
 \begin{figure}[h]
 \includegraphics[width=8cm]{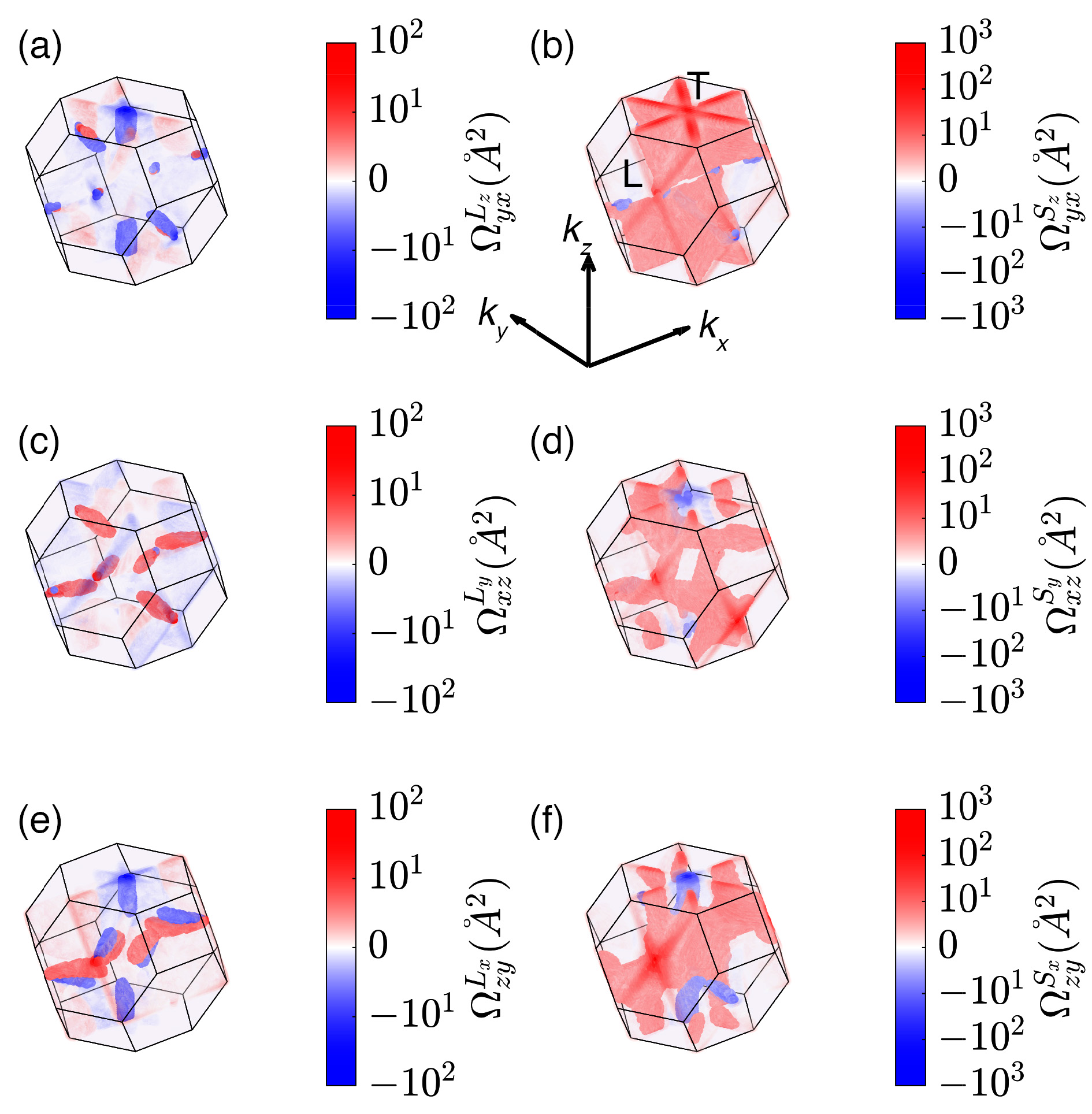}
 \caption{\label{fig:2}  Mapping of (a,c,e) total orbital Berry curvature  ($\Omega^{L_k}_{ji}$)  and (b,d,f) total spin Berry curvature ($\Omega^{S_k}_{ji}$) in the first Brillouin zone. }
 \end{figure}
 %%%%%%%%%%%  Fig:2 %%%%%%%
The intrinsic SHC is usually considered as a pure Fermi sea effect which counts all the occupied bands, see Eq.~\ref{eq:2}. However, in metallic system band pairs with avoided crossings near the Fermi surface often provides the dominant contributions, since spin Berry curvature of bands lying far below the Fermi energy normally are compensated by its counterpart band \cite{guo2008intrinsic}. For Bi semimetal, the Fermi surface consists one hole ellipsoid around T point and three electron ellipsoids around L points \cite{PhysRevX.5.021022}. Figure 2 shows the conventional components of total orbital (spin) Berry curvature, $\Omega^{L_k}_{ji}$ $(\Omega^{S_k}_{ji})$, in the whole first Brillouin zone. We find the major contributions of $\Omega^{L_k}_{ji}$ and $\Omega^{S_k}_{ji}$ mainly locate at those T and L ellipsoids. 
Specifically, the components $\Omega^{L_z}_{yx}$ and $\Omega^{S_z}_{yx}$ whose orbital and spin polarization along the trigonal axis are dominated at T ellipsoid and have opposite sign, see Figs.~\ref{fig:2} (a,b). The components whose polarization along bisectrix $\Omega^{X_y}_{xz}$  and binary $\Omega^{X_x}_{zy}$ axes show dominant contributions at the L ellipsoids, where $\Omega^{L_x}_{zy}$ and $\Omega^{S_x}_{zy}$ clearly show same sign, see Figs.~\ref{fig:2} (e,f). Consequently, anisotropy of SHC in Bi can be simply attributed to the difference between the detailed structure of T and L ellipsoids, while the nearly isotropic SHC within the basal plane indicates spin degree of freedom of the three L ellipsoids is largely independent with their orbital configurations. Conversely, OHC whose polarization within the basal plane still shows a strong anisotropy, reflecting the strong anisotropy of orbital polarization of the three L ellipsoids along the binary $(L_x)$ and bisectrix $(L_y)$ axes. Similarly, huge orbital magnetoresistance of Bismuth \cite{zhu2012field,PhysRevX.5.021022} has been reported due to its strong anisotropy of mobility and band effective mass. Therefore, the distinct anisotropy of the OHC and SHC within the basal plane questions their correlation with the isotropic SOC strength \cite{go2018intrinsic}.

\begin{figure}[h]
 \includegraphics[width=16cm]{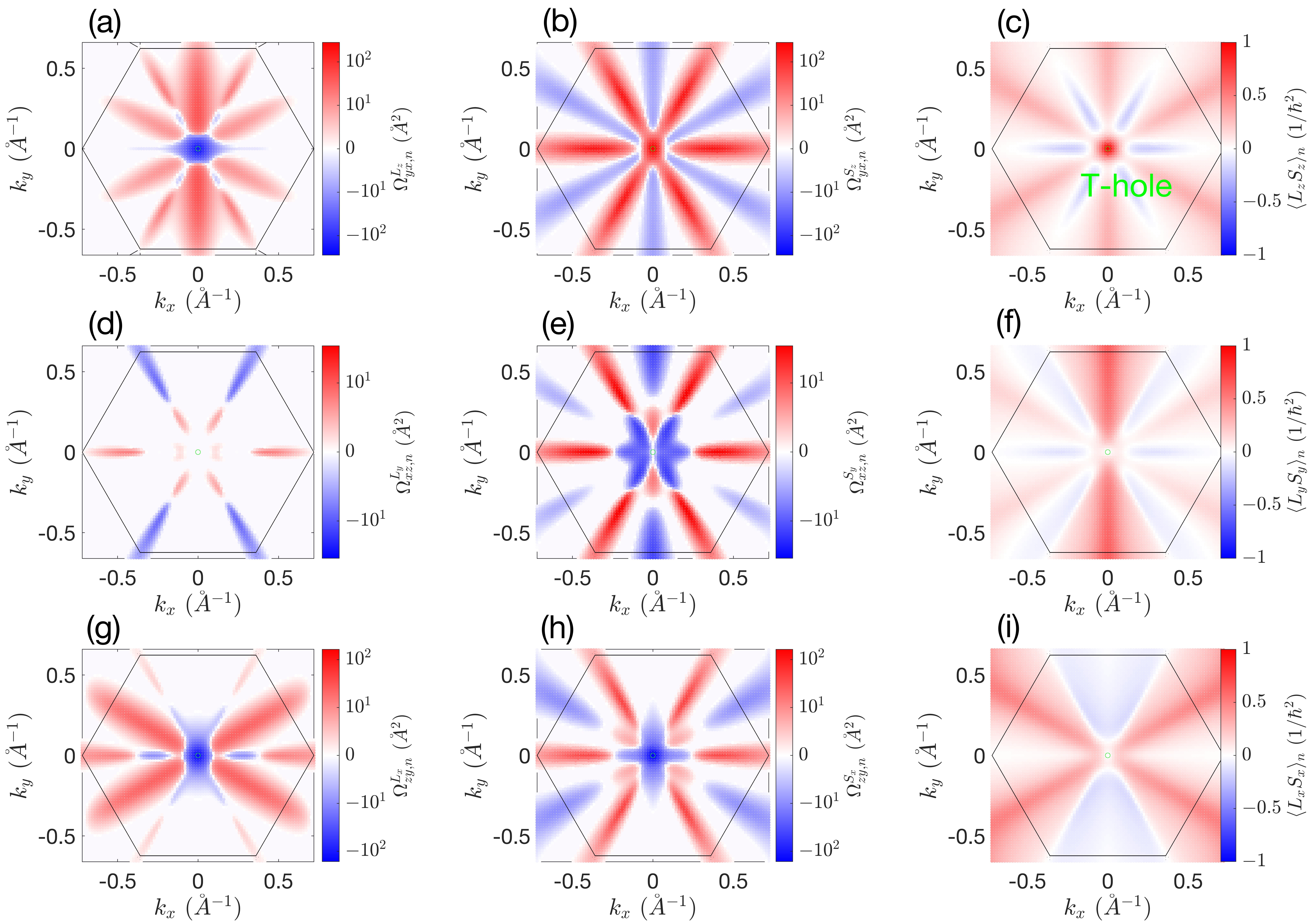}
 \caption{\label{fig:3} Mapping of (a,d,g) orbital Berry curvature ($\Omega^{L_k}_{ji,n}$) and (b,e,h) spin Berry curvature ($\Omega^{S_k}_{ji,n}$) and (c,f,i) partial spin-orbit coupling $\braket{L_k S_k}_n$ of band $n=5$ at top surface of first Brillouin zone center at T. The green circle is the Fermi contour of T-hole pocket. }
 \end{figure}

Figure~\ref{fig:3} presents the orbital and spin Berry curvature $\Omega^{L_k,S_k}_{ji}$ and the partial SOC strength $\braket{L_k S_k}_n$ for the band crossing the Fermi energy $(n=5)$ at the top surface of first Brillouin zone. The Fermi contour of band $5$ shows tiny hole pocket around T point [see Fig.~\ref{fig:3} (c)]. For orbital and spin polarization along the trigonal $L_z,S_z$ axis [see Figs.~\ref{fig:3} (a,b,c)], the partial SOC strength $\braket{L_z S_z}$ present perfectly a three-fold symmetry and $\Omega^{S_z}_{yx}$ qualitatively follows its symmetry, while $\Omega^{L_z}_{yx}$ is, surprisingly, reduced to two-fold. Importantly, at the T point where the partial SOC $\braket{L_z S_z}$ shows strong positive correlation ($\braket{L_z S_z}_n>0$), $\Omega^{L_z}_{yx}$ and  $\Omega^{S_z}_{yx}$ are in opposite sign. Similar feature could also be traced from the cases with  orbital and spin polarization along bisectrix $L_y,S_y$ and binary axes $L_x,S_x$. For example, in the line $k_y=0$ of Figure~\ref{fig:3}(d,e,f), the partial SOC $\braket{L_y S_y}$ shows negative correlation, where $\Omega^{L_y}_{xz}$ and  $\Omega^{S_y}_{xz}$ have same sign; at the T point of Figure~\ref{fig:3}(g,h,i) where partial SOC $\braket{L_x S_x}$ is zero indicating no correlation between spin and orbital, $\Omega^{L_x}_{zy}$ and  $\Omega^{S_x}_{zy}$ coincide with same sign. Thus, it is evident that the orbital and spin Berry curvature are not simply correlated by the SOC strength, especially in the strong SOC regime. 

\subsection{SOC strength and hopping parameter}
 
\begin{figure}[h]
 \includegraphics[width=8cm]{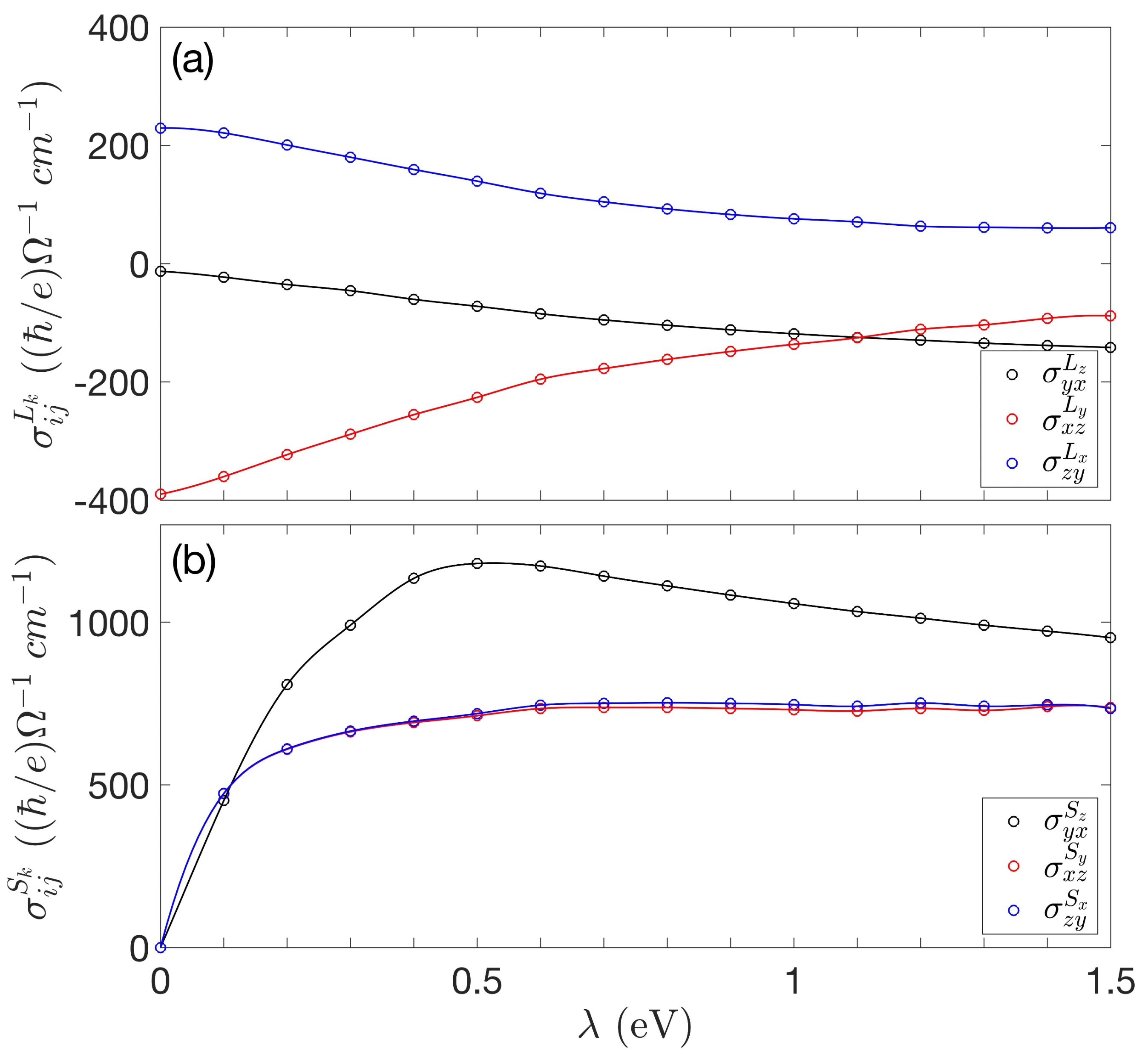}
 \caption{\label{fig:4}  Dependence of orbital Hall conductivity ($\sigma^{L_k}_{ji}$) and spin Hall conductivity ($\sigma^{S_k}_{ji}$) on the strength of spin orbital coupling strength ($\lambda$).  }
 \end{figure}

\begin{figure}[h]
 \includegraphics[width=8cm]{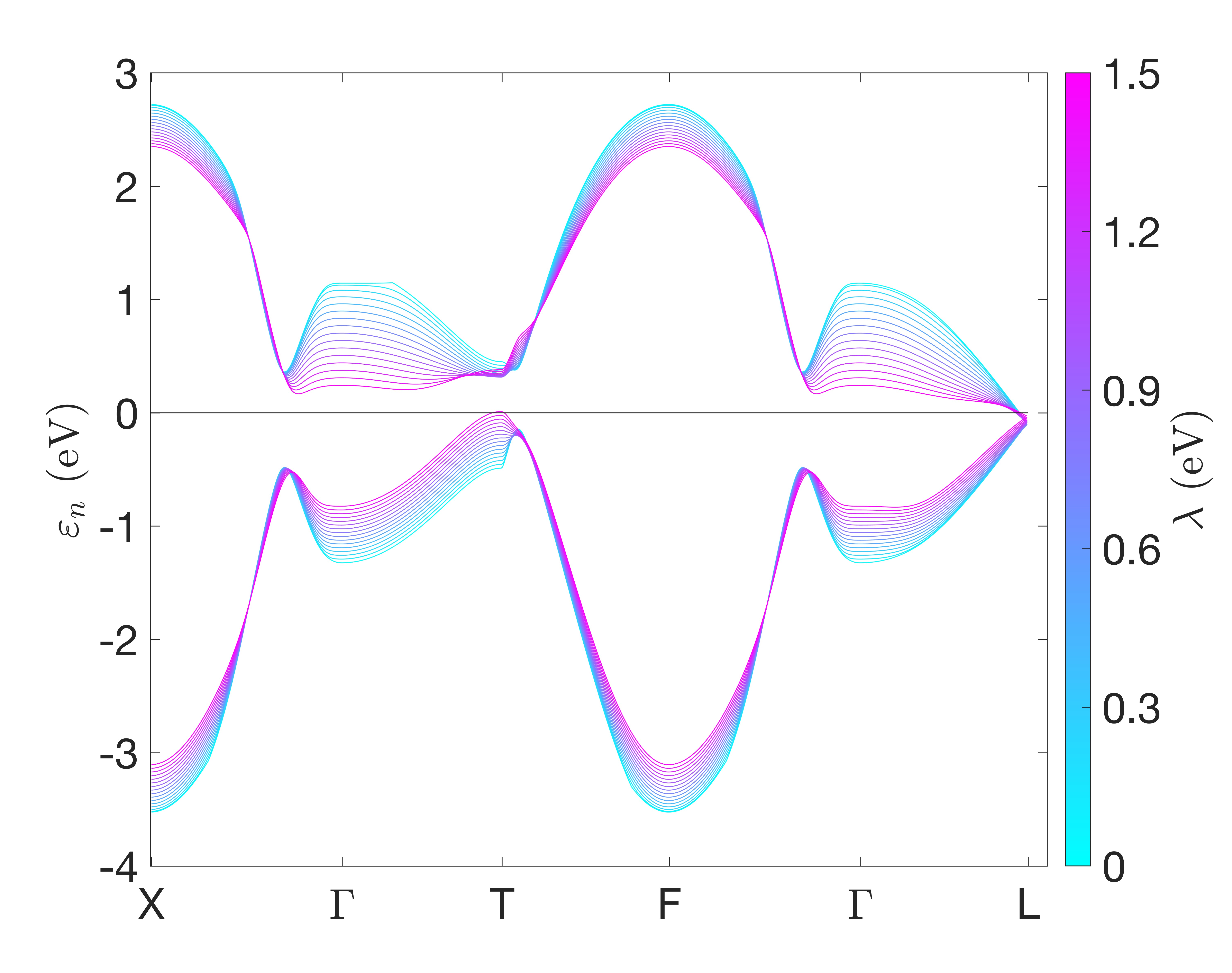}
 \caption{\label{fig:5} Band deformation of lowest conduction band and highest valance band with tuning SOC strength $\lambda$. }
 \end{figure}

Figure~\ref{fig:4} shows the SOC strength $\lambda$ dependence of OHC and SHC. We confirm that SHE vanishes exactly in the absence of SOC while the OHE still survives, indicating that SOC is not the origin of OHE. Conversely, each component of $\sigma^{S_k}_{ji}$ drastically increases with SOC strength up to $0.5$ eV, despite the continuous decrease of $\sigma^{S_z}_{yx}$ with further increasing $\lambda$. Note that the two components with spin polarization in the basal plane $\sigma^{S_y}_{xz},\sigma^{S_x}_{zy}$ mostly coincide during evolution of SOC strength. For the OHE,  $\sigma^{L_y}_{xz}$ and $\sigma^{L_x}_{zy}$ whose orbital polarization within basal plane monotonically decrease with SOC strength, while $\sigma^{L_z}_{yx}$ increases with $\lambda$. We emphasize in the absence of SOC, $\sigma^{L_z}_{yx}$ is nearly zero while $\sigma^{L_y}_{xz}$ and $\sigma^{L_x}_{zy}$ reach their maximum to $458$ and $-780$  $(\hbar/e)  \Omega^{-1} cm^{-1}$, respectively. 
%%%%%%%%%% Fig:3 %%%%%%%%% 
\onecolumngrid\
 \twocolumngrid\
%%%%%%%%%%%  Fig:3 %%%%%%%%
 The opening of SOC strongly affects the  $p$-orbitals hybridization and thus modifies its orbital texture \cite{go2018intrinsic} which suppresses $\sigma^{L_y}_{xz}$ and $\sigma^{L_x}_{zy}$ but enhances the $\sigma^{L_z}_{yx}$ along with the band deformation associated with SOC.  Figure~\ref{fig:5} demonstrates the gradual modification of band dispersion near the Fermi level due to the influence of the strength of SOC which aligns with the gradual variation of OHC. However, all components of SHC monotonically increases along the $p$-orbitals hybridization induced by SOC with a slight decay of $\sigma^{Sz}_{yx}$ until the energy gap is too large to sustain electrons' hopping. The distinct behavior of OHC and SHC with respect to the tuning of SOC strength reiterates their lack of correlation via spin orbit interaction. %Thus, our results imply that the SHE solely relies on SOC.

%%%%%%%%%% Fig:4 %%%%%%%%% 
 %%%%%%%%%%%  Fig:4 %%%%%%%%% 
 
 %%%%%%%%%% Fig:5 %%%%%%%%% 
 %%%%%%%%%%%  Fig:5 %%%%%%%%% 
 
Next we exploit the advantage of the tight-binding to investigate the contribution of each type of orbital hopping and the results are listed in Table~\ref{tab:1}. $\sigma^{L_k,S_k}_{ji}$ are calculated by turning on the hoppings of $s$-$s$, $s$-$p$, and  $p$-$p$ orbitals in velocity operator, while the band structure is kept intact. We find the $p$-$p$ orbital hopping dominates the OHE and SHE in the Bi semimetal, consistent with our analysis on the $p$-orbital spin-orbit hybridization. It is not surprising, since $s$-orbitals are lying far below the Fermi energy. Interesting, if only turning on the $s$-$s$ orbital hopping, the OHE vanishes exactly due to the zero angular momentum of $s$-orbitals, but the SHE still exists. This hypothetical scenario shows the possibility of SHE without companion of OHE despite its tiny size. We argue the non-zero SHE should be attributed to the spin-orbit hybridization, since the band structure still involves SOC.

%%%%%%%%%% Tab:1 %%%%%%%%% 
 \begin{table}[h]
\caption{\label{tab:1} Calculated orbital Hall conductivity $\sigma^{L_k}_{ji}$ and spin Hall conductivity  $\sigma^{S_k}_{ji}$. Tot. represents turning on all hoppings in velocity matrix elements, while $s-s$, $s-p$, and $p-p$ represent turning on only $s$-$s$ hoppings ($V^{\ ,',''}_{ss}$), $s$-$p$ hoppings ($V^{\ ,',''}_{sp}$), and  $p$-$p$ hoppings ($V^{\ ,',''}_{pp}$), respectively. The unit is $(\hbar/e) \Omega^{-1} cm^{-1}$.}
\begin{ruledtabular}
\begin{tabular}{lllllll}
   & $\sigma^{L_z}_{yx}$ & $\sigma^{L_y}_{xz}$ & $\sigma^{L_x}_{zy}$ & $\sigma^{S_z}_{yx}$ & $\sigma^{S_y}_{xz}$ & $\sigma^{S_x}_{zy}$ \\
  \hline
\text{Tot.}  & -283.6 & -176.6 & 121.2 & 952.6 & 737.8  & 735.1\\
$s$-$s$  & 0 & 0 & 0 & 0.05  & 0.08  & 0.09 \\
$s$-$p$  & -0.4  & -6.2 & -2.6 & -7.9 &  -6.2 & -7.7 \\
$p$-$p$  & -196.4  & -230.4 & 265.6 & 1034.3 & 771.9  & 795.4 \\
\end{tabular}
\end{ruledtabular}
\end{table}
%%%%%%%%%% Tab:1 %%%%%%%%% 

\section{Discussion}\label{Sec6:Discussion}

 Though SHE \cite{hirsch1999spin} is historically reported before the formulation of OHE \cite{bernevig2005orbitronics}, SHE is considered as ramification of the latter with the assistance of spin orbit interaction \cite{go2018intrinsic}. Nonetheless, we find the orbital and spin Berry curvature are lack of correlation with SOC strength at Fermi surface. Furthermore, for a hexagonal (rhombohedral) crystal, OHC of Bi with orbital polarization in the basal plane shows large anisotropy but its SHC is approximately isotropic, manifesting the huge distinction between OHE and SHE. By tuning the strength of SOC, OHE continuously changes with band deformation, indicating a continuous change in orbital texture. In contrast, SHE is not solely determined by orbital texture and does not increase monotonically with SOC strength in strong SOC regime. Therefore, in realistic multi-band materials, both OHE and SHE are strongly affected by SOC, but a simple relationship between the two mediated by SOC is debatable.
 
 %Specifically, by tuning the SOC strength, OHE continuously changes with band deformation, suggesting a continuous change of orbital texture, while SHE is not solely determined by orbital texture and is not monotonically increasing with SOC strength.  Thus, in realistic multi-band materials, OHE and SHE are both strongly affected by the SOC but a simple relationship between the two mediated by the SOC is debatable. 

In Bi semimetal, the magnitude of SHC is generally three-fold of OHC, which can be extended to other $p$-orbital materials with gigantic SOC strength for searching large spin current generation. The SHC result is quantitively consistent with measurements on polycrystalline samples \cite{PhysRevB.105.214419}. We also notice the recent experiments on the spin charge conversion of Bismuth semimetal \cite{liang2022anisotropic}, which both reported a huge anisotropy of SHC. However, the totally vanish of SHE along specific crystalline orientation is still intriguing \cite{fukumoto2022observation}. Our result calls for further experimental investigations on anisotropy of orbital (spin) current polarizing in basal plane which might differentiate the two effects due to their distinct anisotropy, e.g., by comparing the spin-charge conversion efficiency with polarization along $[100]_H$ and $[110]_H$ directions. %which might differentiate the two effects due to their distinct anisotropy. 

It should be noted that both orbital and spin current can not be directly measured. The clear evidence of SHE relies on the magneto-optical Kerr effect which probing the spin accumulation at the edges \cite{kato2004observation,PhysRevLett.119.087203}. However, the existing electrical and optical methods are incapable to distinguish the SHE and OHE which both can induce magnetic moment accumulations at edges, if the OHE truly existed. We hope the magnetic circular dichroism experiments detecting the core level electrons \cite{o1994orbital,schattschneider2006detection,verbeeck2010production} can solve this ambiguity.

In conclusion, we demonstrated that in a multi-band system orbital and spin Berry curvature are lack of correlation meditated by SOC at Fermi surface and the OHE and SHE are not correlated in the strong SOC regime. The two effects should be treated as of the equal footing and both are strongly affected by the SOC and orbital hybridization.  We report that in Bi semimetal the intrinsic SHC has comparatively large magnitude and smaller anisotropy, while its intrinsic OHC is small with strong anisotropy. The $p$-orbital materials with strong orbital hybridization and gigantic SOC strength could be seen as promising candidates for realizing large SHE without heavy metals and for studying OHE and SHE comparatively.

\begin{acknowledgments}
%put your acknowledgments here.
The authors would like to thank M. Hayashi for his kind correspondences. This work was supported by Grants-in-Aid for Scientific Research (B) (No. 21H01034) from the Japan Society for the Promotion of Science.
\end{acknowledgments}

%\appendix

%%%%%%%%%%% Appendix A:  FF %%%%%%%%%%%%%%%%

\bibliography{Bi_SHE&OHE}% Produces the bibliography via BibTeX.

\end{document}